\documentclass[apj,numberedappendix]{emulateapj}
\usepackage{apjfonts}

\newcommand{\be}{\begin{eqnarray}}
\newcommand{\ee}{\end{eqnarray}}
\newcommand{\lp}{\left(}
\newcommand{\rp}{\right)}
\newcommand{\lb}{\left[}
\newcommand{\rb}{\right]}
\newcommand{\vrun}{V_{\rm run}}
\newcommand{\rhorun}{\rho_{\rm run}}

\slugcomment{Submitted for publication in The Astrophysical Journal on June 12, 2009.}
\shorttitle{SHOCK BREAKOUT FROM SNe IA}
\shortauthors{PIRO, CHANG, \& WEINBERG}

\begin{document}

\title{Shock Breakout from Type Ia Supernova} 

\author{Anthony L. Piro, Philip Chang\altaffilmark{1}, and Nevin N. Weinberg}

\affil{Astronomy Department and Theoretical Astrophysics Center, University 
of California, Berkeley, CA 94720;\\
tpiro@berkeley.edu,
pchang@astro.berkeley.edu,
nweinberg@astro.berkeley.edu}

\altaffiltext{1}{Canadian Institute for Theoretical Astrophysics, 60 St. George Street, University of Toronto, ON M5S 3H8, Canada; pchang@cita.utoronto.ca}


\begin{abstract}
The mode of explosive burning in Type Ia SNe remains an outstanding problem. It is generally thought to begin as a subsonic deflagration, but this may transition into a supersonic detonation (the DDT). We argue that this transition leads to a breakout shock, which would provide the first unambiguous evidence that DDTs occur. Its main features are a hard X-ray flash ($\sim20\ {\rm keV}$) lasting $\sim10^{-2}\ {\rm s}$ with a total radiated energy of $\sim10^{40}\ {\rm ergs}$, followed by a cooling tail. This creates a distinct feature in the visual light curve, which is separate from the nickel decay. This cooling tail has a maximum absolute visual magnitude of $M_V\approx-9$ to $-10$ at $\approx1\ {\rm day}$, which depends most sensitively on the white dwarf radius at the time of the DDT. As the thermal diffusion wave moves in, the composition of these surface layers may be imprinted as spectral features, which would help to discern between SN Ia progenitor models. Since this feature should accompany every SNe Ia, future deep surveys (e.g., $m=24$) will see it out to a distance of $\approx80\ {\rm Mpc}$, giving a maximum rate of $\sim60\ {\rm yr^{-1}}$. Archival data sets can also be used to study the early rise dictated by the shock heating (at $\approx20\ {\rm days}$ before maximum B-band light). A similar and slightly brighter event may also accompany core bounce during the accretion induced collapse to a neutron star, but with a lower occurrence rate.
\end{abstract}



\section{Introduction}\label{sec:introduction}

   The use of Type Ia supernovae (SNe Ia) as cosmological distance
indicators has brought attention to the theoretical  uncertainties that
remain about these events. Generally it is thought
that they result from the unstable thermonuclear ignition of a C/O
white dwarf (WD). One key question is how the burning
front propagates during the incineration. The consensus is that the
flame begins as a subsonic deflagration \citep{nom76,nom84} to match the
observed nucleosynthesis and light curves \citep{fil97}, but the
later propagation is more uncertain.
Motivated by terrestrial combustion, many have argued for
a delayed detonation transition \citep[DDT,][]{kho91,ww94}.
Although a DDT may be needed to better replicate observations \citep{ple04,liv05},
how and if it happens is uncertain \citep{nw97,nie99,woosley07}.

   Any observational feature that unambiguously demonstrates that a detonation
occurs would be helpful for resolving this uncertainty.
One consequence of a detonation is that it drives a
shock through the WD surface layers, the breakout of which is
a clear signature of the DDT. For Type II supernovae (SNe II),
it was long expected that shock breakout from core-bounce would produce an
X-ray and/or ultra-violet flash \citep{col74,kc78,fal78}.
As the earliest electromagnetic emission available, it is an important
probe of the progenitor star and its circumstellar environment
\citep{mm99}. These X-ray flashes have been observed in
the cases of GRB 060218 \citep{cam06}. which was associated with
the SN Ic 2006aj \citep{maz06} and the X-ray transient XRT 080109,
associated with SN 2008D \citep{sod08}. Indeed, since the SN Ia rate in star forming regions is $\sim1/3$ that of SNe II \citep{man05},
there may be hope that a SN Ia shock breakout
may soon be seen. The question is how bright is such an event and how long does it last

   In this paper, we characterize the light curve expected from an SN Ia
shock breakout.
In \S \ref{sec:shocked}, we present the initial WD profile before shock passage
and investigate when the shock runs away form the detonation using both a numerical,
hydrodynamic simulation and analytic arguments.
We then present equations that describe how the
shock steepens as it propagates through the surface layers ahead of the detonation.
In \S \ref{sec:lightcurve}, we follow the expansion and cooling of the shock-headed envelope using a semi-analytic analysis. We demonstrate
that the entire light curve can be modeled with self-similar solutions.
The breakout produces a sharp, hard X-ray flash with a duration
of $\approx10^{-2}\ {\rm s}$ \citep[also see][]{ims81}. As the shocked envelope expands and cools, the
lightcurve shifts from the X-ray to the UV and eventually to the visual wavebands. In
\S \ref{sec:optical} we show that it reaches a peak absolute magnitude of $M_V=-9$ to $-10$
at $\approx1\ {\rm day}$. We also discuss what depths of the WD are probed by the receding
thermal diffusion wave as a function of time.
In \S \ref{sec:conclusions} we summarize our results and discuss the possibility of
shock breakouts from accretion induced collapse (AIC) to a neutron star.
   

\section{The Dynamics and Thermodynamic Properties of the Shocked Atmosphere}\label{sec:shocked}

We first summarize the properties of the unshocked WD envelope (\S \ref{sec:preshocked}) and then model the detonation to determine how the outward propagating shock steepens as it runs down the background density gradient (\S \ref{sec:runaway}). From these we derive the equations that describe how the envelope responds to the shock's passage (\S \ref{sec:heatingandexpansion}).


\subsection{The Pre-shocked Atmosphere}\label{sec:preshocked}

   The pre-shocked envelope, which we denote with the subscript $0$, is estimated as a constant
flux atmosphere, $F=\sigma_{\rm SB}T_{\rm eff}^4$, and is in radiative equilibrium. We assume a Chandrasekhar mass ($M=1.4M_\odot$) WD with a radius
$R_*\approx(3-6)\times10^8\ {\rm cm}$, which is larger than the pre-ignition
radius due to expansion during the
deflagration phase (in contrast, the pre-ignition radius is
$\approx1.6\times10^8\ {\rm cm}$). Radiative diffusion
gives
\be\label{eq:flux}
	T_{\rm eff}^4 = \frac{16T_0^3g}{3\kappa}\frac{dT_0}{dP_0},
\ee
where we take  a constant opacity, $\kappa=0.2\ {\rm cm^2\ g^{-1}}$, corresponding to
electron scattering in hydrogen deficient material. We take the gravitational acceleration
$g=GM/R_*^2$ to be constant in these shallow, surface layers.
Integrating equation (\ref{eq:flux}), assuming an ideal gas equation of state,
$P_0=\rho_0k_{\rm B}T_0/\mu m_p$, where $\mu$ is the mean molecular weight,
we get the pressure as a function of density,
\be\label{eq:p0}
	P_0(\rho_0) 
	&=& 6.1\times10^{13}\ g_9^{-1/3}T_{\rm eff,5}^{4/3}\rho_0^{4/3}\ {\rm ergs\ cm^{-3}},
\ee
where $g_9=g/10^9\ {\rm cm\ s^{-2}}$, $T_{\rm eff,5}=T_{\rm eff}/10^5\ {\rm K}$, $\rho_0$
is in cgs units, and $\mu=4/3$ (for a helium
dominated composition).

   At sufficiently large depths, the pressure becomes dominated by non-relativistic
electrons, in which case
\be\label{eq:p0deg}
	P_0(\rho_0)= 9.91\times10^{12}(\rho_0/\mu_e)^{5/3}\ {\rm ergs\ cm^{-3}},
\ee
where $\mu_e$ is the mean molecular weight per electron. This switch occurs around
$\rho_0\approx9\times10^3\ {\rm g\ cm^{-3}}$.
These two power laws (eqs. [\ref{eq:p0}] and [\ref{eq:p0deg}])
motivate our use of a polytropic background model in the following sections.


\subsection{Detonation and Shock Runaway}
\label{sec:runaway}

   The energy budget available to the shock is set by where it runs ahead of the detonation,
giving a characteristic velocity and density for the shock, $\vrun$ and $\rhorun$. The strong shock
subsequently steepens in the declining density gradient according to
\be\label{eq:shock velocity}
	V_{0} =\vrun\left(\frac{\rho_0}{\rhorun}\right)^{-\beta},
\ee
with $\beta= 0.1858$ for a radiation pressure dominated shock \citep[][$\gamma=4/3$]{sak60}.

\begin{figure}
\epsscale{1.15}
\plotone{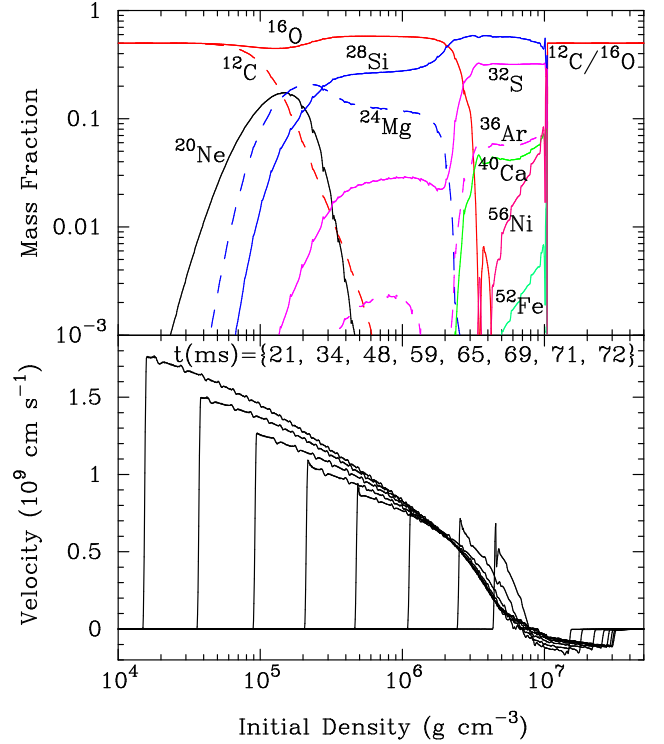}
\caption{The composition ({\it top panel}) and fluid velocity ({\it bottom panel}) as a function of initial WD density.
The detonation is initiated at $10^7\ {\rm g\ cm^{-3}}$, sending a burning wave
into the flammable lower-density region, and a shock wave down into the WD. At
$\approx2\times10^{6}\ {\rm g\ cm^{-3}}$, the steepening shock begins to race ahead of the burning,
as
can be seen by the marked decline in intermediate mass element synthesis.  At this depth, the shock has a speed of $\approx6\times10^8\ {\rm cm\ s^{-1}}$, which sets the initial conditions for the
shock breakout calculation.}
\label{fig:rhov}
\epsscale{1.0}
\end{figure}

   We use a one-dimensional, hydrodynamics code
(described previously in Weinberg \& Bildsten 2007)
to follow the detonation's propagation and transition to a shock, which we show in Figure \ref{fig:rhov}. The background is initialized with a constant flux radiative
atmosphere, which smoothly transitions to a nearly isothermal degenerate star.
A detonation is initiated at $10^7\ {\rm g\ cm^{-3}}$ \citep[near where the DDT is expected,][]{woosley07} by
artificially raising the temperature. Above this depth the composition is a flammable
mixture of equal parts
$^{12}$C and $^{16}$O, and below it is inert material to prevent the detonation from
propagating inward (in accord with the preceding deflagration).
The simulation shows an outward propagating detonation, which
is accompanied by a weak shock sent downward into the WD core.
The detonation propagates until $\approx2\times10^{6}\ {\rm g\ cm^{-3}}$,
at which point the shock runs away. This is most clearly seen in the top panel of Figure \ref{fig:rhov}
where the synthesis of intermediate mass elements falls off in comparison to $^{16}$O
(the depletion of $^{12}$C at these depths is due to residual burning behind the shock).
In the bottom panel, we see that the velocity falls off as a
$V_0\propto \rho_0^{-0.186}$ power law as expected (eq. [\ref{eq:shock velocity}]).
From this plot we determine that the
characteristic numbers at runaway are $\rhorun\approx2\times10^6\ {\rm g\ cm^{-3}}$ and
$\vrun\approx6\times10^8\ {\rm cm\ s^{-1}}$.

   Similar results are obtained by using simple analytic scalings.
As the detonation propagates, it causes a change of pressure
\be
	\frac{p_{\rm det}}{p_0}\approx \frac{E_{\rm nuc}}{E_{\rm int}},
\ee
where $E_{\rm nuc}\approx0.8\ {\rm MeV\ nucl^{-1}}$
is the specific energy release from burning two $^{12}$C and two $^{16}$O to $^{56}$Ni,
$E_{\rm int}=3E_{\rm F}/4\mu_e$
is the internal energy, and
\be
	E_{\rm F}=0.41\rho_6^{1/3}(2/\mu_e)^{1/3}\ {\rm MeV\ nucl^{-1}},
\ee
is the Fermi energy. The speed and temperature of the detonation are then
$T_{\rm det}=(3p_{\rm det}/a)^{1/4}$ and $V_{\rm det}=(p_{\rm det}/p_0)^{1/2}c_s$, respectively, where $c_s$ is the sound speed. At the depths where the electrons are degenerate and relativistic, this gives
a density independent detonation speed of $V_{\rm det}\approx6\times10^8\ {\rm cm\ s^{-1}}$,
consistent with our numerical results.

The burning in the detonation occurs over an ``induction length'' given by
\be
	\lambda = (E_{\rm nuc}/\epsilon)V_{\rm det}
\ee
where $\epsilon$ is the energy generation rate. As the density
decreases, the burning rates does as well, increasing $\lambda$. Once
$\lambda$ is greater than a pressure scaleheight $H=P/\rho g$,
the burning can no longer keep up and the shock races ahead of the burning
front. Using the energy generation rate, $\epsilon$,
for burning $^{12}$C to $^{24}$Mg from Woosley at al. (2004, see their eq. [4]),
this occurs at a critical density of
\be
	\rho_{\rm run}\approx2\times10^6\ g_9^{0.11}{\rm g\ cm^{-3}}.
\ee
The prefactor is robust because of the strong temperature and density
scalings of the energy generation rate. This compares well with our numerical calculation
(see Fig.~\ref{fig:rhov}).


\subsection{Shock Heating and Adiabatic Expansion}\label{sec:heatingandexpansion}

   The jump conditions for a strong shock
(Whitham 1958, his eq.[32]) give the pressure $p_{\rm sh}$, density $\rho_{\rm sh}$, and
velocity $V$ behind the shock in terms of the shock velocity $V_0$,
\be\label{eq:jump conditions}
	p_{\rm sh}= \frac 2 {\gamma + 1} \rho_0 V_0^2,\
	\rho_{\rm sh} = \frac {\gamma + 1}{\gamma- 1} \rho_0,\ V = \frac 2 {\gamma + 1} V_0.
\ee
We next need to understand how these properties change with expansion.

   Consider a shell of mass $\Delta M$, density $\rho_0$, and thickness $\Delta r_0\ll r_0$,
where $r_0\approx R_*$ is the local radius in the WD envelope.
The shell's radius as a function of time is
\be\label{eq:radius}
	r(\rho_0,t) = r_0+V(\rho_0)t\approx R_* + V(\rho_0) t.
\ee
Pressure gradients may increase the velocity $V(\rho_0)$
by as much as a factor of 2 \citep{mm99}, but we find this difference has little effect on our solutions.
By continuity of mass, the shock compresses the shell to a thickness
$\Delta r_{\rm sh}=(\gamma-1)/(\gamma+1)\Delta r_0$.
Subsequently, the thickness increases due to the difference in velocity
between the top and bottom of the shell, giving a thickness as a function
of time
\be\label{eq:thickness}
	\Delta r(\rho_0,t) = \Delta r_{\rm sh} + \Delta V(\rho_0) t,
\ee
where $\Delta V(\rho_0)$ is
\be\label{eq:deltav}
	\Delta V(\rho_0)
	\approx \left.\frac {\partial V}{\partial \rho_0}\right|_{\rho_0}
		\frac{\partial \rho_0}{\partial r_0}\Delta r_0.
\ee
We take the equation of hydrostatic balance,
$dP_0/dr_0 = -\rho_0 g$,
and assume a polytropic equation of state, $P_0 = K\rho_0^{(n+1)/n}$
(where $K$ and $n$ are set by either eq. [\ref{eq:p0}] or [\ref{eq:p0deg}]).
Setting $\Delta r_0 \approx H_0=P_0/\rho_0 g = K\rho_0^{1/n}/g$, where
$H_0$ is the pressure scale height,
we find
\be
	\Delta V(\rho_0) = \frac {n\beta} {n+1} \frac 2 {\gamma + 1}
	\vrun\left(\frac{\rho_0}{\rhorun}\right)^{-\beta},
\ee
which is smaller than $V$ by a constant factor of $n\beta/(n+1)\approx0.14$
(for $n=3$ and $\beta=0.186$).

By mass conservation
\be
	4\pi\rho_0R_*^2 \Delta r_0 = 4\pi \rho(\rho_0,t) r(\rho_0,t)^2 \Delta r(\rho_0,t),
\ee
where $\rho(\rho_0,t)$ is the expanded density as a function of time.
This gives us a relation for the fractional change in density
\be\label{eq:rho(t)}
	\frac{\rho(\rho_0,t)}{\rho_0} = \frac {R_*^2}{r(\rho_0,t)^2}\frac {\Delta r_0}{\Delta r(\rho_0,t)},
\ee
which is a function of only $\rho_0$ and $t$.

From the shock jump conditions (eq.[\ref{eq:jump conditions}]), we find the pressure in the shocked
and expanded shell,
\be\label{eq:psh}
	p _{\rm sh}(\rho_0) = \frac 2 {\gamma + 1} \rho_0 V_0^2
	= \frac 2 {\gamma + 1} \rhorun \vrun^2 \left(\frac {\rho_0}{\rhorun}\right)^{1-2\beta}.
\ee
We assume that the pressure decreases adiabatically as the shell expands, i.e., $P\propto\rho^\gamma$, so that
\be\label{eq:p(t)}
	p(\rho_0,t)&=&p_{\rm sh}(\rho_0)\lb \frac{\rho(\rho_0,t)}{\rho_0}\rb^{\gamma}
	\nonumber
	\\
	&=& \frac 2 {\gamma + 1} \rhorun \vrun^2\left(\frac {\rho_0}{\rhorun}\right)^{1-2\beta}
	\lb \frac{\rho(\rho_0,t)}{\rho_0}\rb^{\gamma}.
\ee
Although the adiabatic approximation is good, it is in general not exactly true, an issue that we discuss in \S \ref{sec:case1}.
By combining equations (\ref{eq:rho(t)}) and (\ref{eq:p(t)}), we solve
for the pressure of the expanded shell with initial density $\rho_0$ at any time $t$.


\section{The Light Curve from Shock Breakout and Shock-heated Cooling}\label{sec:lightcurve}

\subsection{The Prompt Breakout Flash}

   Photons in the radiative shock stream out once $\tau\lesssim c/V$, where
$\tau$ is the optical depth. Substituting $V\rightarrow \Gamma V$, where $\Gamma=[1-(V/c)^2]^{-1/2}$ (the shock is mildly relativistic at breakout), we find a breakout depth of
\be\label{eq:rhobr}
	\rho_{\rm 0,br} &=& \rhorun\lb \frac{\gamma+1}{2}\frac{\Gamma c}{\vrun}\frac{g}{\kappa K\rhorun^{1+1/n}}\rb^{1/(1+1/n-\beta)}
	\nonumber
	\\
	&\approx&10^{-3}(g_9/K_{13.8})^{0.87}V_9^{-0.87}\rho_6^{-0.16}(\Gamma/1.7)^{0.87}\ {\rm g\ cm^{-3}},
	\nonumber
	\\
\ee
where $K_{13.8}=K/6\times10^{13}\ {\rm cgs}$, $V_9=\vrun/10^9\ {\rm cm\ s^{-1}}$, and $\rho_6=\rhorun/10^6\ {\rm g\ cm^{-3}}$ (using typical values from Fig.~\ref{fig:rhov}).

   This initial streaming of photons gives rise to the prompt breakout flash. The energy density available is $E=3p_{\rm sh}$, and using equation (\ref{eq:psh}) for $p_{\rm sh}$, we estimate a total energy
budget at a depth $\rho_{\rm 0,br}$ of
\be\label{eq:eflash}
	E_{\rm flash}&=& 4\pi R_*^2H_0\times3p_{\rm sh}
	\nonumber
	\\
	&\approx&4\times10^{40}\ (g_9/K_{13.8})^{-0.16}V_9^{1.0}\rho_6^{0.22}R_{8.5}^2\ {\rm ergs},
\ee
where $R_{8.5}=R_*/3\times10^8\ {\rm cm}$, and associated temperature
\be
	T_{\rm flash} &=& (3p_{\rm sh}/a)^{1/4}
	\nonumber
	\\
	&\approx& 2\times10^8\ (g_9/K_{13.8})^{0.14}V_9^{0.32}\rho_6^{0.068}\ {\rm K}.
\ee
This temperature is a lower limit, since non-LTE effects (which we ignore to simplify our analysis) will only harden the spectrum \citep{kat09}. We further discuss the limitations of this assumption in \S \ref{sec:lte}. These numbers are consistent with the findings of \citet{ims81}, who predicted a hard ($\approx20\ {\rm keV}$) X-ray flash associated with WD detonation. The main difference is that we are quantifying this flash in terms of the DDT picture. The timescale for the energy release will be dominated light travel effects ($R_*/c\approx10^{-2}\ {\rm s}$), which we include in our calculations in \S  \ref{sec:cooling}. Also, the DDT will not occur everywhere at once \citep[for example, as found in the simulations by][]{rop07}, which further smears out the breakout flash.

\subsection{Cooling of the Ejected Shock-Heated Envelope}\label{sec:cooling}

Following breakout, a thermal diffusion wave begins propagating back into the shock-heated envelope,
releasing photons that will be seen as the shock breakout flash.
At any given depth in the shock-heated envelope, the energy density is $E=3p$
and leaks out of the envelope on a thermal diffusion
timescale, which at a depth $\rho_0$ and a time $t$, is
\be\label{eq:tdiff}
	t_{\rm diff}(\rho_0,t) = \frac{\kappa}{c} \rho(\rho_0,t) \lb\Delta r(\rho_0,t)\rb^2.
\ee
Since $t_{\rm diff}$ sets the timescale when an observer sees down to a certain depth
in the shock heated envelope, we set $t_{\rm diff}(\rho_0,t)=t$ to solve for $\rho_0(t)$,
using the prescription for $\rho(\rho_0,t)$ and $\Delta r(\rho_0,t)$ from
\S \ref{sec:heatingandexpansion}.
Once we have $\rho_0(t)$, we can solve for any other property of the envelope as a
function of time. In particular, the luminosity of the expanding and cooling envelope is
\be\label{eq:luminousity}
	L(t) = \frac {4\pi r(t)^2 E(t)c} {\tau(t)},
\ee
where
\be
	\tau(t) = \kappa\Delta r(t)\rho(t)
\ee
is the optical depth of the diffusion layer.


\begin{figure}
\epsscale{1.2}
\plotone{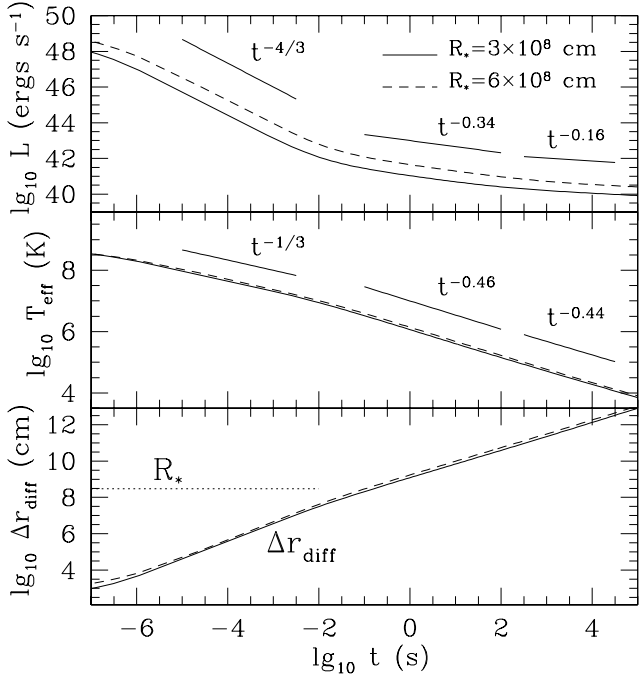}
\caption{Solutions for the cooling, shock-heated WD envelope with $\vrun=10^9\ {\rm cm\ s^{-1}}$ and $\rhorun=10^6\ {\rm g\ cm^{-3}}$.
The top and middle panels show the time-dependent luminosity
and effective temperature from the moment of shock breakout until $\approx 1\ {\rm day}$.
The bottom panel shows the thermal diffusion depth $\Delta r_{\rm diff}\equiv\Delta r(t=t_{\rm diff})$.
The solid lines ({\it dashed lines}) are for an initial WD radius of $R_*=3\times10^8\ {\rm cm}$ ($6\times10^8\ {\rm cm}$).
In the bottom panel, the horizontal dotted line denotes $R_*=3\times10^8\ {\rm cm}$.}
\label{fig:lightcurve}
\epsscale{1.0}
\end{figure}

\begin{figure}
\epsscale{1.2}
\plotone{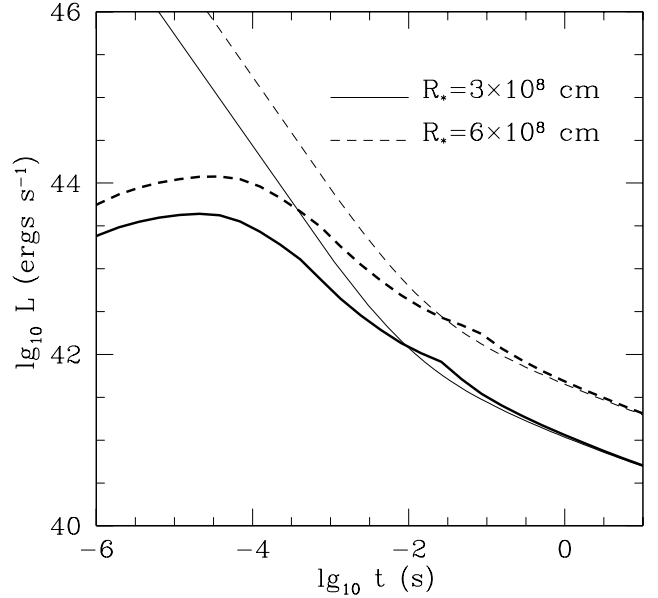}
\caption{The early time light curve, including time travel effects across the WD surface ({\it thick lines}) in comparison to the one-dimensional solutions plotted in Figure \ref{fig:lightcurve} ({\it thin lines}). Light travel effects decrease the observed luminosity at early times when light from the entire surface has not yet reached the observer.}
\label{fig:lighttravel}
\epsscale{1.0}
\end{figure}

   The general solution of $L(t)$ is non-algebraic, but easily solved numerically.
In Figure \ref{fig:lightcurve} we plot $L(t)$
({\it top panel}), the effective temperature ({\it middle panel})
\be\label{eq:teff}
	T_{\rm eff}(t) = \lb \frac{L(t)}{4\pi r(t)^2\sigma_{\rm SB}}\rb^{1/4},
\ee
and the thermal diffusion depth $\Delta r_{\rm diff}\equiv\Delta r(t=t_{\rm diff})$ ({\it bottom panel}). The luminosity
decreases as a broken power law in time. The initial decline is steeper, with a $L\propto t^{-4/3}$ power law. In practice, this power law and the very high initial luminosities are an artifact of our one-dimensional treatment, and will not be seen in observations. Just as for the breakout flash, light travel effects and not simultaneous DDT ignition will smear out the light curve over $R_*/c\approx10^{-2}\ {\rm s}$ or more. To better quantify this effect, we plot the light curves altered by light travel effects in Figure \ref{fig:lighttravel}. These are calculated according to
\be
	L(t) = \int_0^{\pi/2}L[t_r(t,\cos\theta)]d\cos\theta,
\ee
where $t_r=t-r(t_r)(1-\cos\theta)/c$ is the retarded time of photons emitted at a angle $\theta$ with respect to the observer. Note that $r(t_r)$ must be evaluated at the retarded time because the radius was smaller in the past
The robust feature to take away is that $\sim10^{40}\ {\rm ergs}$ of energy is released during this initial peak (consistent with eq. [\ref{eq:eflash}]).
In the second stage, the luminosity decreases more slowly,
from $10^{42}\ {\rm ergs\ s^{-1}}$ to $10^{40}\ {\rm ergs\ s^{-1}}$ over
$\approx 1\ {\rm day}$.
The effective temperature evolution covers a wide range of wavelengths, going from
X-rays to ultraviolet to visual over the course of $\approx 1\ {\rm day}$.

   The power law behavior of $L(t)$ can be understood analytically by considering the expansion and
cooling in certain limiting cases. To emphasize this, we have plotted $R_*=3\times10^8\ {\rm cm}$
in the bottom panel of Figure \ref{fig:lightcurve} ({\it dotted line}).
A comparison with $L(t)$ clearly
shows that the luminosity break occurs when $\Delta r_{\rm diff}$ has expanded to a thickness $\approx R_*$.
Motivated by this, we consider the time-dependent behavior in two different limits:
\begin{enumerate}
\item At early times, $t \ll R_*/V$, the envelope has barely moved and $r(t)\approx R_*$.
By equation (\ref{eq:rho(t)}), we see that $\rho/\rho_0\propto t^{-1}$ since
the radial position of the shell remains constant and the thickness increases as $\Delta r\propto t$.
\item At late times, $t \gg R_*/V$, both $r$ and $\Delta r$ increase like $t$, so that
$\rho/\rho_0 \propto t^{-3}$.
\end{enumerate}
We next derive the analytic self-similar solutions in each limit.

\subsection{Case 1: $t\ll R_*/V$}\label{sec:case1}

   In this limit the expansion is plane-parallel so that the
radius is basically fixed at $r(t)\approx R_*$, whereas the thickness
has expanded by $\Delta r(t)\approx \Delta V t$.
Since $\rho(t)\propto t^{-1}$, we find that the thermal diffusion depth is independent of
time when we set the condition $t_{\rm diff}=t$, and we find
\be\label{eq:rho_diff2}
	\rho_{\rm 0,diff} &=& \frac{c}{\Delta V}\lp\kappa H_0\rp^{-1}
	\nonumber
	\\
	&\approx&4\times10^{-3}\ (g_9/K_{13.8})^{0.87}V_9^{-0.87}\rho_6^{-0.16}
	\ {\rm g\ cm^{-3}}.
\ee
The physics expressed by equation (\ref{eq:rho_diff2}) is that the thermal diffusion wave sits at a fixed depth because the column of material sitting above it is fixed.
The optical depth is proportional to column, and thus also fixed,
\be
	\tau_{\rm diff} = \kappa \Delta V t \rho(t)=\kappa H_0\rho_0
	\approx7(g_9/K_{13})^{0.87}V_9^{-0.87}\rho_6^{-0.16}.
	\nonumber
	\\
\ee
This gives a luminosity of
\be
	L &=& \frac{4\pi R_*^2E(t)c}{\tau}
	\nonumber
	\\
	&\approx&9\times10^{41}\ (g_9/K_{13.8})^{-0.34}V_9^{0.72}\rho_6^{0.13}R_{8.5}^2t_{-2}^{-4/3} {\rm ergs\ s^{-1}},
	\nonumber
	\\
\ee
where $t_{-2}=t/10^{-2}\ {\rm s}$ and
\be
	T_{\rm eff} \approx1\times10^{7}\ (g_9/K_{13.8})^{-0.086}V_9^{0.18}\rho_6^{0.033}t_{-2}^{-1/3}{\rm K}
\ee
is the effective temperature.

 Since in this limit, the same mass shell is always doing the radiating (as indicated by a fixed $\rho_{\rm 0,diff}$ in eq. [\ref{eq:rho_diff2}]), it cannot be
 evolving adiabatically (thus our assumption in eq. [\ref{eq:p(t)}] is not strictly valid). In a more detailed
study we solve the entropy equation in the plane parallel limit, including
the radiative loss term (Chang, Matzner, \& Piro, in preparation). We find
that the cooling is only marginally faster when we include radiative losses,
with a logarithmic time dependence.

   The plane-parallel limit applies until $Vt\approx R_*$. Setting $\rho_{\rm 0,diff}=\rho_0$ into equations (\ref{eq:shock velocity}) and (\ref{eq:jump conditions}), we find this occurs at
  \be
	t\approx1\times10^{-2}\ (g_9/K_{13.8})^{0.16}V_9^{-1.2}\rho_6^{-0.22}R_{8.5}\ {\rm s}
\ee
at which point we must begin to account for the radial expansion.

\subsection{Case 2: $t \gg R_*/V$}

The layer has now moved an appreciable distance from
the star, so that $r(t)\approx Vt$ \citep[as addressed in][]{che92}. Solving for the diffusion
depth in this limit
\be\label{eq:rhodiff3}
	\rho_{\rm 0,diff} &=& \rhorun\lb \frac{n+1}{n\beta}\frac{2}{\gamma+1}
	\frac{g}{\kappa K\rhorun^{1+1/n}}
	\frac{\vrun}{c}
	\lp\frac{ct}{R_*}\rp^2
	\rb^{1/(1+\beta+1/n)},
	\nonumber
	\\
	&\approx&2\ (g_9/K_{13.8})^{0.66}V_9^{0.66}\rho_6^{0.12}R_{8.5}^{-1.3}t^{1.3}\ {\rm g\ cm^{-3}}.
\ee
The optical depth is
\be
	\tau &=& \frac{c}{\Delta V}=\frac{n+1}{n\beta}\frac{\gamma+1}{2}\frac{c}{\vrun}
	\lp \frac{\rho_0}{\rhorun}\rp^\beta
	\nonumber
	\\
	&=&21\ (g_9/K_{13.8})^{0.12}V_9^{-0.88}\rho_6^{-0.16}R_{8.5}^{-0.25}t^{0.25}.
\ee
This gives a luminosity of
\be
	L(t)\approx 3\times10^{41}\ (g_9/K_{13.8})^{-0.50}V_9^{1.8}\rho_6^{0.42}R_{8.5}^{1.0}t^{-0.34}\ {\rm ergs\ s^{-1}},
	\nonumber
	\\
\ee
which has a power law time dependence, $L\propto t^{-0.34}$, consistent with what \citet{che92} found in the same regime. The effective temperature is
\be
	T_{\rm eff}(t) \approx 1\times10^6\ (g_9/K_{13.8})^{-0.065}V_9^{0.019}\rho_6^{0.0035}R_{8.5}^{0.13}t^{-0.46}\ {\rm K}.
	\nonumber
	\\
\ee
Since the surface area is increasing and the diffusion wave is moving into higher temperature material, the evolution is shallower than in the plane-parallel case.

\subsection{The Exposure of Originally Degenerate Material}

   At sufficiently late times, the diffusion wave moves into material that was originally
degenerate before being hit by the shock.  Setting $\rho_{\rm 0,diff}$ in equation
(\ref{eq:rhodiff3}) equal to $\approx9\times10^3\ {\rm g\ cm^{-3}}$ (\S \ref{sec:preshocked}),
this occurs at $\approx700\ {\rm s}$. The dynamics are essentially the same as that of \S 3.2, but now $n=3/2$ instead of $n=3$.  The diffusion wave now moves through the layer more slowly, with $\rho_{\rm diff,0}\propto t^{1.1}$, which gives a luminosity,
\be
	L(t) \approx 2\times10^{40}\ (g_9/K_{13})^{-0.41}V_9^{1.9}\rho_6^{0.36}R_{8.5}^{0.83}t_4^{-0.16}\ {\rm ergs\ s^{-1}},
	\nonumber
	\\
\ee
and effective temperature,
\be
	T_{\rm eff}(t) = 2\times10^4\ (g_9/K_{13})^{-0.058}V_9^{0.030}\rho_6^{0.0058}R_{8.5}^{0.11}t_4^{-0.44}\ {\rm K},
	\nonumber
	\\
\ee
where $K_{13}=K/10^{13}\ {\rm cgs}$ and $t_4=t/10^4\ {\rm s}$.
The initially degenerate material causes the light curve to flatten and the luminosity remains at $\sim10^{40}\ {\rm ergs\ s^{-1}}$ well until a day after the initial shock breakout (as is shown in Fig. \ref{fig:lightcurve}).


\subsection{Assumption of Local Thermodynamic Equilibrium}\label{sec:lte}

   In our derivations we have assumed that the photons and electrons are thermally equilibrated
throughout the shock passage and subsequent expansion.  This was done for the sake of attaining concrete results and simple analytic expressions, but it begs the question of how accurate this is.

   After the initial shock passage, the envelope is heated and the electron and photons can reach the same temperature via Comptonization on a timescale $t_{\rm Comp} \sim (m_ec^2/k_{\rm B}T(\rho_0))/(\kappa\rho_0 c)$, where $m_e$ is the electron mass and $T(\rho_0)$ is solved from equation (\ref{eq:psh}). This timescale must be less than the local expansion timescale, $t_{\rm exp}\sim H_0/\Delta V(\rho_0)$, otherwise significant expansion occurs before equilibrium is reached. For the condition $t_{\rm Comp}\lesssim t_{\rm exp}$, we find that the density must be greater than
\be
	\rho_0\gtrsim 3\times10^{-3}\  (g_9/K_{13})^{0.59}V_9^{0.030}\rho_6^{0.0055}\ {\rm g\ cm^{-2}}.
\ee
This is not much greater than the density of the shock breakout (eq. [\ref{eq:rhobr}]), so we conclude that at least initially all but the very outer envelope reaches equilibrium.

   Subsequently, expansion and adiabatic cooling can drive the photons and electron back out of thermal equilibrium. And in fact, the timescale for thermalization via Comptonization is always shorter than the thermal diffusion time at the diffusion depth. Since the photons and electrons were coupled in the past, we don't expect the spectrum to be altered too greatly, but this should be quantified by more detailed calculations in the future.

\section{Optical Light Curve}\label{sec:optical}
   
\begin{figure}
\epsscale{1.2}
\plotone{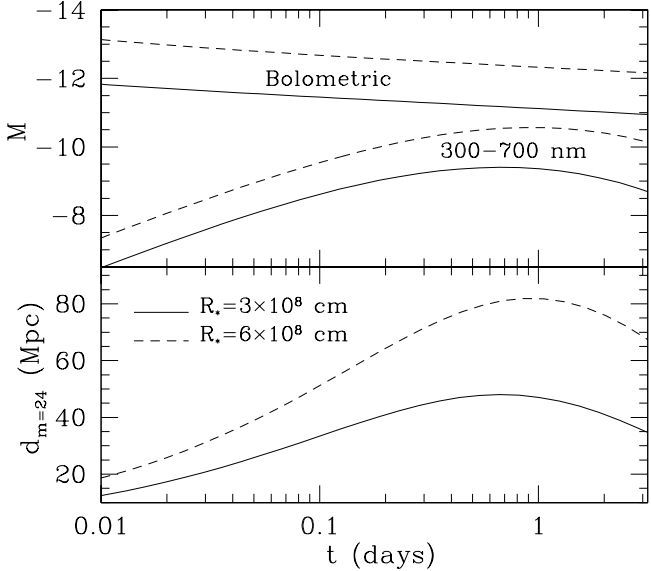}
\caption{The top panel plots the absolute magnitude of the expanding and cooling shock heated envelope. The upper curves show the bolometric luminosity, and the lower ones show the absolute magnitude in the wavelength range of $300-700\ {\rm nm}$. The solid lines ({\it dashed lines}) are for $R_*=3\times10^8\ {\rm cm}$ ($6\times10^8\ {\rm cm}$). The bottom panel shows the distance out to which an $m=24$ limited optical survey could see such an event. }
\label{fig:filter}
\epsscale{1.0}
\end{figure}

   In Figure \ref{fig:filter} we show the bolometric and optical ($300-700$ nm) absolute magnitude for two different progenitor radii, assuming that the WD emits as a blackbody. In the bottom panel we plot the distance out to which such an optical event can be seen for an $m=24$ limited exposure. The optical peaks at around $\approx1\ {\rm day}$ following shock breakout. The observability depends sensitively on the WD radius at the time of shock breakout.
   
   Although we have assumed a blackbody emitter for these light curve estimates, the atmosphere will in fact be scattering dominated. This hardens the spectrum, making the cooling tail dimmer in the optical. Also, for the wavelengths that dominate at $\approx1\ {\rm day}$, metals may play an important role in setting the opacity, especially since they will experience some recombination for the temperatures at this time. More sophisticated spectral modeling is needed in order to predict the precise shape of the light curve.

      In Figure \ref{fig:filter} we have focused on the light curve due to from the cooling of the shock-heated envelope, but eventually this will be overtaken by the nickel decay. Whether or not the shock-heated cooling can be seen above the rising light curve of nickel decay depends on the power of the early time nickel decay and the WD radius at the time of the DDT. If we extrapolate the typical $L\propto t^2$ law found empirically \citep{con06},
or the exponential luminosity function of \citet{arn82}, back to early times, they would go right through our optical light curves, indicating that nickel decay will be comparable to our optical cooling luminosity at $\approx1\ {\rm day}$. Since such extrapolations to early times are uncertain, early sampling of the SN Ia light curves is needed to better address this issue. As our calculations show, detecting the signature of a SN Ia at such early times is possible with current and future surveys. Using a SN Ia rate of
$2.93\times10^{-5}\ {\rm yr^{-1}\ Mpc^{-3}}$ \citep{dil08}, we estimate that $\approx10-60$ SNe Ia can be observed at these early times each year.

\begin{figure}
\epsscale{1.2}
\plotone{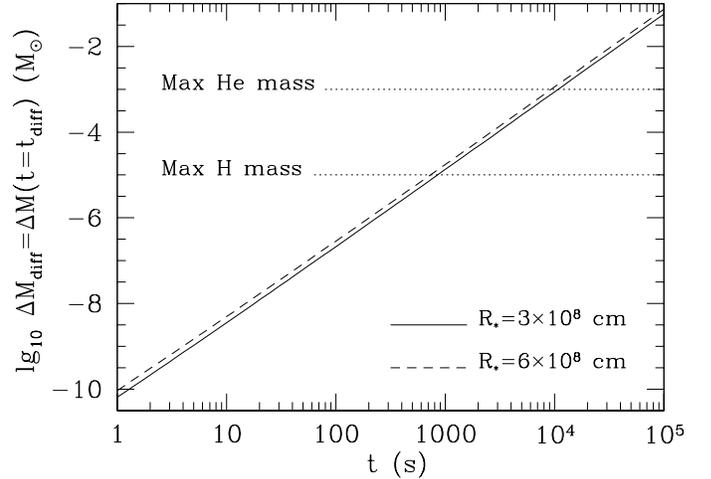}
\caption{The depth of the thermal diffusion wave during cooling following shock-heating of the WD envelope as a function of time. Also marked by dotted lines are estimates for the maximum mass helium shell ($\approx10^{-3}\ M_\odot$, Iben \& Tutukov 1989; Shen \& Bildsten 2009b) and the maximum mass hydrogen shell ($\approx10^{-5}\ M_\odot$; Shen \& Bildsten 2009a) that can survive on a $1.4\ M_\odot$ WD without igniting. This shows what layers of the WD surface may be probed by the thermal diffusion wave to help discern between possible SN Ia progenitors.}
\label{fig:mass}
\epsscale{1.0}
\end{figure}

   Detailed spectral modeling of the shock-heated cooling may also help discern between SN Ia progenitor models. In Figure \ref{fig:mass}, we plot the thermal diffusion depth in units of mass. This shows which mass shells of the progenitor WD are being probed by the cooling wave as a function of time. Depending on the progenitor model in question, different compositions are expected in these surface layers, and once the thermal diffusion wave is below them, these elements may imprint their presence as spectral features. As a comparison, we plot dotted lines denoting the characteristic maximum thickness of a hydrogen shell  ($\approx10^{-5}\ M_\odot$; Shen \& Bildsten 2009a) or helium shell ($\approx10^{-3}\ M_\odot$, Iben \& Tutukov 1989; Shen \& Bildsten 2009b), as expected to be present in the single degenerate scenario for SNe Ia. 

\section{Summary and Conclusions}\label{sec:conclusions}

   We have calculated the cooling of a shock heated WD envelope due to a DDT during a SN Ia. The general features of the resulting flash are a short ($\sim10^{-2}\ {\rm s}$), hard X-ray flash with a total energy of $\sim10^{40}\ {\rm ergs}$. The flash is followed by a cooling tail as the thermal diffusion wave travels back into the expanding envelope. The light curve transitions from the X-rays to the ultraviolet, and eventually, visual wavebands, reaching
a peak absolute magnitude of $M_V=-9$ to $-10$ at $\approx1\ {\rm day}$.
   Whether or not this emission is observable as a distinct component in the SNe Ia optical light curve depends on the WD radius at the time of DDT and the nature of the early-time nickel decay. Nevertheless, the detection (or lack thereof) of emission during the first $\sim20\ {\rm days}$ before peak would be an important constraint on the occurrence of DDTs and on the amount of expansion during the deflagration phase.
   
   A similar shock breakout event may also be associated with the AIC of a WD to a neutron star. In fact, \citet{tan01} and \citet{des07} considered whether the breakout shock from core bounce in this case would be a gamma-ray burst progenitor, concluding the answer is no. But this does not preclude a less powerful, but nevertheless interesting, shock breakout as we have described here. The energy is greater in the AIC case than the SN Ia case, because it is powered by the gravitational binding energy of the WD (mediated by neutrinos). Using a typical energy of $10^{50}\ {\rm ergs}$ deposited into $0.1\ M_\odot$ \citep[as found by][]{fry99}, this gives $\vrun\approx10^9\ {\rm cm\ s^{-1}}$, but with a much larger $\rhorun\approx10^7\ {\rm g\ cm^{-3}}$. This provides a late time light curve that is brighter by a factor of $\approx2.5$, making it easier to observe. However, the AIC rate is much lower than the SNe Ia rate \citep[$\sim1\%$,][]{yl98}, and the shock breakout may be confined along the poles \citep{des06,des07}, both of which will make it difficult to catch an AIC shock breakout.


\acknowledgments

We thank Lars Bildsten, Chris Matzner, and Eliot Quataert for helpful discussions.
This work was supported 
by the Theoretical Astrophysics Center at UC Berkeley.

\end{document}